\documentclass[12pt]{article}

\usepackage{graphicx}
\usepackage{epstopdf}

\begin{document}
\begin{center}
\begin{Large}

{\bf Hartree-Fock approximation for non-Coulomb interactions in three and two-dimensional systems}
\end{Large}
\vspace{1.5cm}

\begin{large}
{\bf Vlad-Mihai Ene, Ilinca Lianu and Ioan Grosu }
\end{large}
\vspace{0.5cm}

{\bf Department of Physics, "Babes-Bolyai" University, 400084
Cluj-Napoca, Romania}
\end{center}
\vspace{1cm}

We analyzed the Hartree-Fock approximation for an electron system. The interaction between particles is modeled by a non-Coulombian potential. We analyzed both the three-dimensional
and two-dimensional systems. We obtained accurate analytical results for the particle energy, the particle velocity, the ground state energy of the system as well as the momentum
dependent density of states. The previous classical results for the Coulombian case were reobtained as particular cases.
\vspace{1cm}

\vspace{1cm}

\begin{large}
{\bf I. Introduction}
\end{large}
\vspace{0.5cm}

 A quantum-mechanical description of the electron motion in a solid requires the many electrons Hamiltonian that contains the kinetic and the interaction contribution. For electrons in a solid the interaction contribution correspond to the potential due to the nuclei and the two particle Coulomb interaction between electrons. In the independent electron approximation one consider that each electron moves in an average external field created by all the other electrons. When one utilizes a product wave-function of the $N$ electrons and the average potential due to all the other electrons one obtains the Hartree approximation [1,2]. The Hartree potential depends on the electron density which itself depends on the single particle wave-functions. From here the Hartree potential needs to be self-consistently determined. The Hartree energy can easily determined considering the motion of a single particle, characterized by the creation operator $c^{\dagger}_{k,\sigma}$, in an environment where the interactions are described by the potential $V_{\vec{q}}$. Using the equation of motion method [3], the motion of the single particle characterized by $k$-momentum and $\sigma$-spin, ($\vec{k},\sigma$), is coupled with the motion of two particles and a hole, described by a $c^{\dagger}_{1}c^{\dagger}_{2}c_{3}$ term. This term leads to the Hartree energy if we keep only the term with $\vec{q}=0$ (i.e. neglecting the density fluctuations) and replace the product $c^{\dagger}c$ with its ground state expectation value. The Hartree-Fock approximation (HFA) goes beyond Hartree approximation taking into account the antisymmetric character of the many-fermions wave functions, incorporating the concept of exchange interaction [4,5]. For non-interacting electrons the antisymmetric wave function is the well-known Slater determinant [2]. The usual approximation is to consider the ions as an uniform positive charge background that compensate the negative charge of the electrons (jellium model). For such uniform systems the HFA reduces to a first order calculation. With the equation of motion method the HFA result for a uniform system is obtained using the linearized result for the trilinear term $c^{\dagger}_{1}c^{\dagger}_{2}c_{3}\rightarrow c^{\dagger}_{1}\left<c^{\dagger}_{2}c_{3}\right>-c^{\dagger}_{2}\left<c^{\dagger}_{1}c_{3}\right>$. On the other hand, the interactions between electrons can lead to instabilities in the system, which are distorsion instabilities of the Fermi surface [6,7]. For a two-dimensional uniform system, considering a central interaction potential, Quintanilla et.al.[8] shows the possibility of a quantum phase transition, of first or of second order, depending on the form of the interaction potential. The importance of the HFA for a two-dimensional jellium system was discussed in [9] in connection with the field of semiconductor quantum dot technology. The temperature effect on exchange Hartree-Fock energy of the two-dimensional electron gas was discussed by Hong and Mahan [10] in connection with trapped electrons in semiconductor interfaces. On the other hand, the systems of many particles in which the interactions are non-Coulombian are of both theoretical and experimental interest [11-17]. For such systems, the Fermi liquid character can be destroyed by long-range interactions, in the dimension $1<d<2$, as was pointed out in Ref.[18].
 Classical particle systems with super-long range type interactions can present special properties, as shown in Ref.[19]. Thus, these (non-integrable) systems can lead to divergences
in the expression of energy, and for these systems the statistical ensembles can be inequivalent. In the case of long-range interacting quantum systems a series of important experimental results were obtained [20], these interactions having a strong influence on the critical behavior (e.g. the susceptibility) of the systems. For bosonic systems, the super-long range type interactions also lead to the modification of the energy spectrum with effects in the properties of e.g. ultracold dipolar systems and superfluid $^{4}He$ [21].
 In this paper we analyze the uniform electron gas with non-Coulomb interactions of the form $r^{-\eta}$. We calculate the particle energy, the Hartree-Fock energy of the system, the particle velocity, and the density of states. The analysis is done for both three-dimensional and two-dimensional systems, assuming the stability of the systems. The difficult problem of stability remains to be analyzed in a future paper. For the three-dimensional system as well as for the two-dimensional one the results for the Coulomb case are reobtained considering $\eta=1$.

\vspace{0.5cm}

\begin{large}
	{\bf II. Model}
\end{large}
\vspace{0.5cm}

{\bf{ A. The three-dimensional system}}
\vspace{0.5cm}

First we will analyze the three-dimensional (3D) case with the Fourier transform of the interacting potential given by:
$$V_{\vec{p}-\vec{k}}=\int d^{3}r\;V(r) e^{-i(\vec{p}-\vec{k})\vec{r}}\eqno{(1)}$$
where $V(r)$ is the interacting potential, and we will use the geometry from Fig.1
\begin{figure}[h]
		\centering
    \includegraphics[width=7cm]{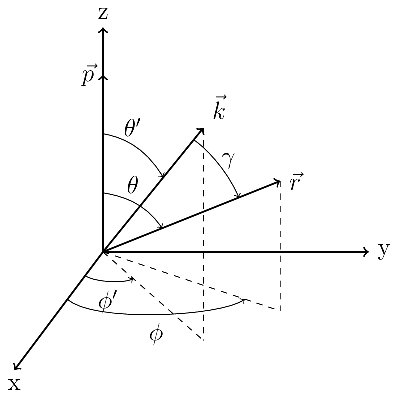}
    \caption{Geometry for $3D$ integration}
    \label{fig1}
\end{figure}
From this, eq.(1) becomes:
$$V_{\vec{p}-\vec{k}}=\int d^{3}r\;V(r) e^{-i p r\cos\theta} e^{i k r \cos\gamma}\eqno{(2)}$$
Using the plane-wave expansion:
$$e^{i k r \cos\alpha}=\sum_{l=0}^{\infty}i^{l} (2l+1) P_{l}(\cos\alpha)j_{l}(k r)\eqno{(3)}$$
where $P_{l}(x)$ are Legendre polynomials and $j_{l}(x)$ are spherical Bessel functions of the first kind, and the following identities [22]:
$$P_{l'}(\cos\gamma)=P_{l'}(\cos\theta)P_{l'}(\cos\theta ')
+ 2\sum_{m=1}^{l'}\frac{(l'-m)!}{(l'+m)!} P_{l'}^{m}(\cos\theta)P_{l'}^{m}(cos\theta ') \cos[m(\varphi-\varphi ')]\eqno{(4)}$$
($P_{l}^{m}(x)$ - is the associated Legendre polynomial) and:
$$\int_{0}^{\pi}d\theta \sin\theta\; P_{l}(\cos\theta)\;P_{l'}(\cos\theta)=\frac{2}{2l+1}\; \delta_{l,l'}\eqno{(5)}$$
($\delta_{l,l'}$ - is the Kronecker delta), one obtains:
$$V_{\vec{p}-\vec{k}}=\sum_{l=0}^{\infty} V_{l}(p,k)\;P_{l}(\cos\theta ')\eqno{(6)}$$
with:
$$V_{l}(p,k)=4\pi (2l+1)\int_{0}^{\infty}dr\;r^{2} V(r)\;j_{l}(p r)j_{l}(k r)\eqno{(7)}$$
The Hartree-Fock energy of an electron, in a homogeneous system, is given by ($\hbar=1$) [4,5]:
$$\varepsilon_{p}=\frac{p^{2}}{2m}-\int\frac{d^{3}k}{(2\pi)^{3}}\;V_{\vec{p}-\vec{k}}\eqno{(8)}$$
or:
$$\varepsilon_{p}=\frac{p^{2}}{2m}-\frac{1}{(2\pi)^{2}}\int_{0}^{p_{F}}\int_{0}^{\pi}dk\;k^{2}\sin\theta' d\theta' \sum_{l=0}^{\infty} V_{l}(p,k)\;P_{l}(\cos\theta ')\eqno{(9)}$$
With the use of the integral:
$$\int_{-1}^{1}dx\; x^{m} P_{n}(x)=0\eqno{(10)}$$
for $m<n$, the electron energy becomes:
$$\varepsilon_{p}=\frac{p^{2}}{2m}-\frac{1}{2\pi^{2}}\int_{0}^{p_{F}}dk\;k^{2} V_{0}(p,k)\eqno{(11)}$$
where $p_{F}$ - is the Fermi momentum. With (7), eq.(11) is rewritten as:
$$\varepsilon_{p}=\frac{p^{2}}{2m}-\frac{2}{\pi}\int_{0}^{p_{F}}dk\;k^{2}\int_{0}^{\infty}dr\;r^{2}V(r)\;j_{0}(p r)j_{0}(k r)\eqno{(12)}$$
Using the formula that connects the spherical Bessel function with the Bessel function of the first kind:
$$j_{\nu}(z)=\sqrt{\frac{\pi}{2z}}\; J_{\nu+\frac{1}{2}}(z)\eqno{(13)}$$
the $\varepsilon_{p}$ energy will be:
$$\varepsilon_{p}=\frac{p^{2}}{2m}-\frac{1}{\sqrt{p}}\int_{0}^{\infty}dr\;r\;V(r) J_{\frac{1}{2}}(p r)\int_{0}^{p_{F}}dk\;k^{3/2} J_{\frac{1}{2}}(k r)\eqno{(14)}$$
The last integral in eq.(14) is easy evaluated to obtain:
$$\varepsilon_{p}=\frac{p^{2}}{2m}-p_{F}\sqrt{\frac{p_{F}}{p}}\int_{0}^{\infty}dr\;V(r) J_{\frac{1}{2}}(p r) J_{\frac{3}{2}}(p_{F}r)\eqno{(15)}$$
Now we will evaluate $\varepsilon_{p}$ for the non-Coulomb interaction described by the potential:
$$V(r)=\frac{A}{r^{\eta}}\eqno{(16)}$$
Here, $A$ is a coefficient whose dimensionality is linked to the number $\eta$, in order to preserve the dimensionality of the potential energy. $\eta$ is a number that
models the non-Coulomb character of the interaction. For $0<\eta<1$ the interaction is of the super long-range type, and for $\eta>1$ it is of the sub-Coulomb type.
Usually, the systems with long-range interactions are defined as the systems where the two-particle potential at large distances decays as $r^{-\eta}$ with $\eta\le d$,
where $d$ is the space dimension, with the Fourier transform of the interaction $V_{\vec{q}}\sim |q|^{\eta-d}$, for $d=3$ and $d=2$. Using eq.(16),
the notation $y=p/p_{F}$, and the new variable $z=p_{F}r$, the $\varepsilon_{p}$ energy becomes:
$$\varepsilon_{p}=\frac{p^{2}}{2m}-A p_{F}^{\eta}\;\sqrt{\frac{1}{y}}\int_{0}^{\infty}dz\;z^{-\eta} J_{\frac{1}{2}}(y z) J_{\frac{3}{2}}(z)\eqno{(17)}$$
and using standard integrals [23] the final result will be:
$$\varepsilon_{p}=\frac{p^{2}}{2m}-A p_{F}^{\eta}
\left\{\begin{array}{lll}
\frac{\Gamma\left(\frac{3-\eta}{2}\right)}{2^{\eta}\Gamma\left(\frac{1}{2}\right)\Gamma\left(1+\frac{\eta}{2}\right)}F\left(\frac{3-\eta}{2}, -\frac{\eta}{2} ; \frac{3}{2} ; y^{2}\right) & \mbox{; $y<1$} \\
\frac{\Gamma(\eta)\Gamma\left(\frac{3-\eta}{2}\right)}{2^{\eta}\Gamma\left(\frac{\eta}{2}\right)\Gamma\left(1+\frac{\eta}{2}\right)\Gamma\left(\frac{3+\eta}{2}\right)} & \mbox{; $y=1$} \\
\frac{\Gamma\left(\frac{3-\eta}{2}\right)}{2^{\eta}y^{3-\eta} \Gamma\left(\frac{5}{2}\right)\Gamma\left(\frac{\eta}{2}\right)}F\left(\frac{3-\eta}{2}, 1-\frac{\eta}{2} ; \frac{5}{2} ; \frac{1}{y^{2}}\right) & \mbox{; $y>1$}
\end{array}\right.\eqno{(18)}$$
for $0<\eta<3$. Here, $\Gamma(z)$ is the Gamma function, and $F(\alpha, \beta ; \gamma ; z)$ is the hypergeometric function.
In Fig.2 we plot, qualitatively, the momentum dependence of the $\varepsilon_{p}$ energy for both cases, $\eta<1$ and $\eta>1$.
\begin{figure}[h]
		\centering
    \includegraphics[width=14cm]{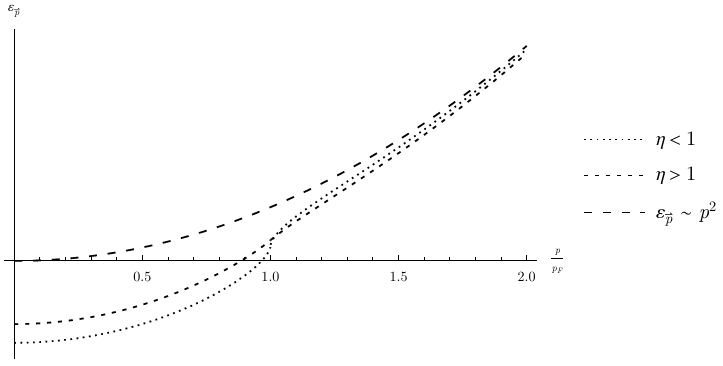}
    \caption{Qualitative dependence of the particle energy, for $\eta<1$ and for $\eta>1$}
    \label{fig2}
\end{figure}
For small values of the particle's momentum the negative exchange energy part dominates. This is a consequence of the Pauli exclusion principle, leading to the well-known
exchange hole space around an electron. The situation is similar to the Coulomb case, the differences being of the quantitative nature.
In the following we will approximate the
$\varepsilon_{p}$ energy for $p$ close to the Fermi momentum $p_{F}$. For the case $p\rightarrow p_{F}; p<p_{F}$, using:
$$F(\alpha, \beta; \gamma; z)=\frac{\Gamma(\gamma)\Gamma(\gamma-\alpha-\beta)}{\Gamma(\gamma-\alpha)\Gamma(\gamma-\beta)} F(\alpha, \beta; \alpha+\beta-\gamma+1; 1-z)+ $$
$$+(1-z)^{\gamma-\alpha-\beta} \frac{\Gamma(\gamma)\Gamma(\alpha+\beta-\gamma)}{\Gamma(\alpha)\Gamma(\beta)} F(\gamma-\alpha, \gamma-\beta; \gamma-\alpha-\beta+1; 1-z)\eqno{(19)}$$
and series expansion for the hypergeometric function we obtain:
$$\varepsilon_{p}\simeq\frac{p_{F}^{2}}{2m}-A p_{F}^{\eta}\frac{\Gamma\left(\frac{3-\eta}{2}\right)}{2^{\eta}\Gamma\left(1+\frac{\eta}{2}\right)}\times $$
$$\times \left\{\frac{\Gamma(\eta)}{\Gamma\left(\frac{\eta}{2}\right)\Gamma\left(\frac{3+\eta}{2}\right)}{\cal{F}}_{1}+
\left(1-y^{2}\right)^{\eta} \frac{\Gamma(-\eta)}{\Gamma\left(\frac{3-\eta}{2}\right)\Gamma\left(-\frac{\eta}{2}\right)}{\cal{F}}_{2}\right\}\eqno{(20)}$$
with:
$${\cal{F}}_{1,2}\simeq 1\mp \frac{\eta (3\mp \eta)}{4 (1\mp \eta)}\left(1-y^{2}\right)\eqno{(21)}$$
The case $y=1$ ($p=p_{F}$) is easily obtained because ${\cal{F}}_{1}={\cal{F}}_{2}=1$, and $\left(1-y^{2}\right)^{\eta}=0$. For the case $p\rightarrow p_{F}$, $p>p_{F}$, in a similar way,
one obtains:
$$\varepsilon_{p}\simeq\frac{p_{F}^{2}}{2m}-A p_{F}^{\eta}\frac{\Gamma\left(\frac{3-\eta}{2}\right)}{2^{\eta}y^{3-\eta} \Gamma\left(\frac{\eta}{2}\right)}\times $$
$$\times \left\{\frac{\Gamma(\eta)}{\Gamma\left(1+\frac{\eta}{2}\right)\Gamma\left(\frac{3+\eta}{2}\right)}{\cal{F}}_{3}+
\left(1-\frac{1}{y^{2}}\right)^{\eta} \frac{\Gamma(-\eta)}{\Gamma\left(\frac{3-\eta}{2}\right)\Gamma\left(1-\frac{\eta}{2}\right)}{\cal{F}}_{4}\right\}\eqno{(22)}$$
where:
$${\cal{F}}_{3,4}\simeq 1+ \frac{(2\mp \eta) (3\mp \eta)}{4 (1\mp \eta)}\left(1-\frac{1}{y^{2}}\right)\eqno{(23)}$$
In order to reobtain the result for the Coulomb case we will use eq.(18) with $A=e^{2}$ and $\eta=1$. In this case we obtain:
$$\varepsilon_{p}=\frac{p^{2}}{2m}-\frac{2 e^{2} p_{F}}{\pi}
\left\{\begin{array}{lll}
F\left(1, -\frac{1}{2} ; \frac{3}{2} ; y^{2}\right) & \mbox{; $y<1$} \\
\frac{1}{2} & \mbox{; $y=1$} \\
\frac{1}{3 y^{2}}\;F\left(1, \frac{1}{2} ; \frac{5}{2} ; \frac{1}{y^{2}}\right) & \mbox{; $y>1$}
\end{array}\right.\eqno{(24)}$$
Using here the following identities [24]:
$$F\left(1, -\frac{1}{2} ; \frac{3}{2} ; y^{2}\right)=\frac{1}{2}+\frac{1-y^{2}}{4 y} \ln\left(\frac{1+y}{1-y}\right)\eqno{(25)}$$
and:
$$F\left(1, \frac{1}{2} ; \frac{5}{2} ; x^{2}\right)=\frac{3}{2 x^{2}}\left[1-\frac{1-x^{2}}{2 x}\ln\left(\frac{1+x}{1-x}\right)\right]\eqno{(26)}$$
the final result is the well known formula:
$$\varepsilon_{p}=\frac{p^{2}}{2m}-\frac{2 e^{2} p_{F}}{\pi}\left[\frac{1}{2}+\frac{1-y^{2}}{4 y}\ln\left|\frac{1+y}{1-y}\right|\right]\eqno{(27)}$$
Now we will evaluate the ground state Hartree-Fock energy of a system of $N$ particles, given by:
$$E_{HF}=E_{HF}^{(1)}+E_{HF}^{(2)}\eqno{(28)}$$
where:
$$E_{HF}^{(1)}=\frac{3}{5}\;\frac{p_{F}^{2}}{2m}\;N\eqno{(29)}$$
is the contribution of the free system, and the exchange contribution is calculated as:
$$E_{HF}^{(2)}=-\frac{A p_{F}^{\eta+3}\Gamma\left(\frac{3-\eta}{2}\right)\; I_{\eta}}{2^{\eta+1}\pi^{2}\Gamma\left(\frac{3}{2}\right)\Gamma\left(1+\frac{\eta}{2}\right)}\eqno{(30)}$$
where:
$$I_{\eta}=\int_{0}^{1}dy\;y^{2} F\left(\frac{3-\eta}{2}, -\frac{\eta}{2}; \frac{3}{2}; y^{2}\right)\eqno{(31)}$$
For the Coulomb interaction $I_{\eta=1}=1/4$, and:
$$E_{HF}^{(2)}=-\frac{e^{2}p_{F}^{4}}{4\pi^{3}}\eqno{(32)}$$
For the non-Coulomb case, using the new variable $x=y^{2}$ and the following integral:
$$\int_{0}^{1}dx\;x^{\gamma-1} (1-x)^{\rho-1} F(\alpha, \beta; \gamma; x)=\frac{\Gamma(\gamma)\Gamma(\rho)\Gamma(\gamma+\rho-\alpha-\beta)}
{\Gamma(\gamma+\rho-\alpha)\Gamma(\gamma+\rho-\beta)}\eqno{(33)}$$
we get for the scaled Hartree-Fock energy the following result:
$$\frac{E_{HF}}{N}=\frac{3}{5}\;\frac{p_{F}^{2}}{2m}-A\left(\frac{p_{F}}{2}\right)^{\eta}\frac{12\; \Gamma\left(\frac{3-\eta}{2}\right)\Gamma(\eta)}
{\eta (\eta+1)(\eta+3)\Gamma\left(\frac{1+\eta}{2}\right)\left[\Gamma\left(\frac{\eta}{2}\right)\right]^{2}}\eqno{(34)}$$
The velocity of the particle is obtained taking the derivative of $\varepsilon_{p}$ with respect to $p$. For the case $y=p/p_{F} <1$, using the first line from (18), one obtains:
$$v_{p}=\frac{p}{m}+\frac{2}{3}\; A p_{F}^{\eta-2}\frac{\eta \Gamma\left(\frac{5-\eta}{2}\right)}{2^{\eta} \Gamma\left(\frac{3}{2}\right)\Gamma\left(1+\frac{\eta}{2}\right)}\;p\;
F\left(\frac{5-\eta}{2}, 1-\frac{\eta}{2}; \frac{5}{2}; \frac{p^{2}}{p_{F}^{2}}\right)\eqno{(35)}$$
and for $y>1$ the velocity becomes:
$$v_{p}=\frac{p}{m}+2 A p_{F}^{\eta-1}\frac{\Gamma\left(\frac{5-\eta}{2}\right)}{2^{\eta} \Gamma\left(\frac{5}{2}\right)\Gamma\left(\frac{\eta}{2}\right)}
\left(\frac{p_{F}}{p}\right)^{4-\eta}\times $$
$$\times\left\{F\left(\frac{3-\eta}{2}, 1-\frac{\eta}{2}; \frac{5}{2}; \frac{p_{F}^{2}}{p^{2}}\right)+\frac{2}{5}\left(1-\frac{\eta}{2}\right)\left(\frac{p_{F}}{p}\right)^{2}
F\left(\frac{5-\eta}{2}, 2-\frac{\eta}{2}; \frac{7}{2}; \frac{p_{F}^{2}}{p^{2}}\right)\right\}\eqno{(36)}$$
The Coulomb case, $A=e^{2}$ and $\eta=1$, is obtained using the following identities:
$$F(a, b; c; x)=\frac{c-1}{a-1}\; x^{-1}\left[F(a-1, b; c-1; x)-F(a-1, b-1; c-1; x)\right]\eqno{(37)}$$
$$F(1, b; c; x)=\frac{c-1}{b-1} (1-x)^{-1}-\frac{c-b}{b-1} (1-x)^{-1} F(1, b-1; c; x)\eqno{(38)}$$
and eqs.(25-26). The velocity will be:
$$v_{p}=\frac{p}{m}+\frac{e^{2}}{\pi}\;\frac{1}{y}\left[\frac{1+y^{2}}{2 y}\ln\left|\frac{1+y}{1-y}\right|-1\right]\eqno{(39)}$$
Using the determined velocities one can evaluate the density of states which, in the three-dimensional case, is [4]:
$$N(p)=\frac{p^{2}}{\pi^{2} \left|v_{p}\right|}\eqno{(40)}$$
Here we will be interested in the momentum dependence of the density of states near the Fermi momentum $p_{F}$.

$\bullet$ For the case $p\rightarrow p_{F}$, $p<p_{F}$, using eq.(19)
and the series of the hypergeometric function, we obtain:
$$v_{p\rightarrow p_{F}}\simeq\frac{p_{F}}{m}+\frac{2}{3}\; A p_{F}^{\eta-1}\frac{\eta \Gamma\left(\frac{5-\eta}{2}\right)}{2^{\eta} \Gamma\left(\frac{3}{2}\right)\Gamma\left(1+\frac{\eta}{2}\right)}\times $$
$$\times\left\{\frac{\Gamma\left(\frac{5}{2}\right)\Gamma(-1+\eta)}{\Gamma\left(\frac{\eta}{2}\right)\Gamma\left(\frac{3+\eta}{2}\right)}{\cal{Q}}_{1}+
\left(1-y^{2}\right)^{-1+\eta}\frac{\Gamma\left(\frac{5}{2}\right)\Gamma(1-\eta)}{\Gamma\left(\frac{5-\eta}{2}\right)\Gamma\left(1-\frac{\eta}{2}\right)}{\cal{Q}}_{2}\right\}\eqno{(41)}$$
where:
$${\cal{Q}}_{1}=F\left(\frac{5-\eta}{2}, 1-\frac{\eta}{2}; 2-\eta; 1-y^{2}\right)\simeq 1+\frac{5-\eta}{4}\left(1-y^{2}\right)\eqno{(42)}$$
and:
$${\cal{Q}}_{2}=F\left(\frac{\eta}{2}, \frac{3+\eta}{2}; \eta; 1-y^{2}\right)\simeq 1+\frac{3+\eta}{4}\left(1-y^{2}\right)\eqno{(43)}$$
At the Fermi momentum $p=p_{F}$, and ${\cal{Q}}_{1}={\cal{Q}}_{2}=1$. Here we distinguish two cases:

a.) Case: $\eta>1$, when $\left(1-y^{2}\right)^{-1+\eta}=0$. In this case:
$$v_{p=p_{F}}=\frac{p_{F}}{2}\left[\frac{2}{m}+A\left(\frac{p_{F}}{2}\right)^{\eta-2} {\cal{M}}(\eta)\right]\eqno{(44)}$$
with:
$${\cal{M}}(\eta)=\frac{\Gamma\left(\frac{5-\eta}{2}\right)\Gamma(-1+\eta)}{\left[\Gamma\left(\frac{\eta}{2}\right)\right]^{2}\Gamma\left(\frac{3+\eta}{2}\right)}\eqno{(45)}$$
The density of states at Fermi momentum has a finite (non-zero) value given by:
$$N(p_{F})=\frac{2 p_{F}}{\pi^{2}\left[\frac{2}{m}+A\left(\frac{p_{F}}{2}\right)^{\eta-2} {\cal{M}}(\eta)\right]}\eqno{(46)}$$

b.) Case: $\eta<1$. In this case the term containing $\left(1-y^{2}\right)^{-1+\eta}$ is dominant in the expression of velocity. Taking this term into account and using the complements
formula for the Gamma function, we get ($\eta\neq 0$):
$$v_{p\rightarrow p_{F}}\simeq\frac{A\Gamma(1-\eta)}{\pi}\left(\frac{p_{F}}{2}\right)^{\eta-1}\left(1-\frac{p^{2}}{p_{F}^{2}}\right)^{-1+\eta}\sin\left(\frac{\pi\eta}{2}\right)\eqno{(47)}$$
and the density of states:
$$N\left(p\rightarrow p_{F}\right)\simeq\frac{2^{\eta-1}}{\pi A\Gamma(1-\eta)\left|\sin\left(\frac{\pi\eta}{2}\right)\right|}\; p_{F}^{3-\eta}\left(1-\frac{p^{2}}{p_{F}^{2}}\right)^{1-\eta}
\eqno{(48)}$$

$\bullet$ For the case $p\rightarrow p_{F}$, $p>p_{F}$, following similar steps, the velocity close to the Fermi momentum will be:
$$v_{p\rightarrow p_{F}}\simeq\frac{p_{F}}{m}+2A p_{F}^{\eta-1}\frac{\Gamma\left(\frac{5-\eta}{2}\right)}{2^{\eta}\Gamma\left(\frac{5}{2}\right)\Gamma\left(\frac{\eta}{2}\right)}\times$$
$$\times\left\{\frac{\Gamma\left(\frac{5}{2}\right)\Gamma(\eta)}{\Gamma\left(1+\frac{\eta}{2}\right)\Gamma\left(\frac{3+\eta}{2}\right)}{\cal{F}}_{3}+
\left(1-\frac{1}{y^{2}}\right)^{\eta}\frac{\Gamma\left(\frac{5}{2}\right)\Gamma(-\eta)}{\Gamma\left(\frac{3-\eta}{2}\right)\Gamma\left(1-\frac{\eta}{2}\right)}{\cal{F}}_{4}+\right.$$
$$\left.+\frac{2}{5}\left(1-\frac{\eta}{2}\right)\left[\frac{\Gamma\left(\frac{7}{2}\right)\Gamma(-1+\eta)}{\Gamma\left(1+\frac{\eta}{2}\right)\Gamma\left(\frac{3+\eta}{2}\right)}{\cal{S}}_{1}+
\left(1-\frac{1}{y^{2}}\right)^{-1+\eta}\frac{\Gamma\left(\frac{7}{2}\right)\Gamma(1-\eta)}{\Gamma\left(\frac{5-\eta}{2}\right)\Gamma\left(2-\frac{\eta}{2}\right)}{\cal{S}}_{2}\right]\right\}
\eqno{(49)}$$
where:
$${\cal{S}}_{1}\simeq 1+\frac{(4-\eta)(5-\eta)}{4 (2-\eta)}\left(1-\frac{1}{y^{2}}\right)\eqno{(50)}$$
and:
$${\cal{S}}_{2}\simeq 1+\frac{(2+\eta)(3+\eta)}{4 \eta}\left(1-\frac{1}{y^{2}}\right)\eqno{(51)}$$
Here, at the Fermi momentum, ${\cal{F}}_{3}={\cal{F}}_{4}={\cal{S}}_{1}={\cal{S}}_{2}=1$, $(1-1/y^{2})^{\eta}=0$, for $\eta>0$, and $(1-1/y^{2})^{-1+\eta}=0$, for $\eta>1$, and diverges
for $\eta<1$.

c.) For the case $\eta>1$ we reobtain the result given by eq.(46) which shows that at $p=p_{F}$ the density of states is non-zero and is continuous.

d.) For the case $\eta<1$ taking again the dominant contribution from the particle velocity, the density of states close to the Fermi momentum will be:
$$N\left(p\rightarrow p_{F}\right)\simeq\frac{2^{\eta-1}}{\pi A\Gamma(1-\eta)\left|\sin\left(\frac{\pi\eta}{2}\right)\right|}\; p_{F}^{3-\eta}\left(1-\frac{p_{F}^{2}}{p^{2}}\right)^{1-\eta}
\eqno{(52)}$$
In Fig.3 we give a qualitative plot of the density of states for $\eta<1$ and for $\eta>1$.
\begin{figure}[h]
		\centering
    \includegraphics[width=14cm]{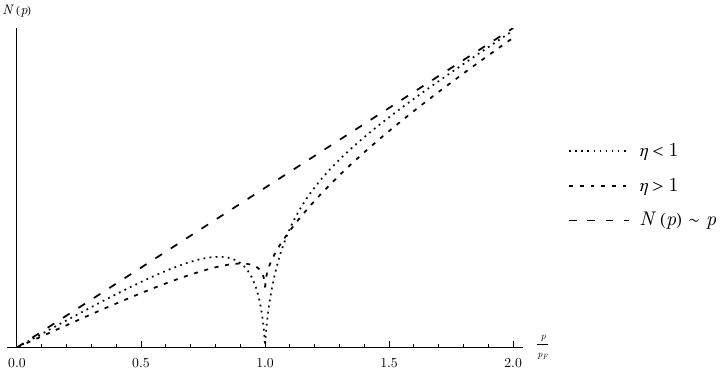}
    \caption{The $3D$ density of states for $\eta<1$ and for $\eta>1$}
    \label{fig3}
\end{figure}
In the case of super-long range interaction ($\eta<1$) the density of states tends to
zero when $p\rightarrow p_{F}$, having a power function type dependence. In the sub-Coulomb case ($\eta>1$) the density of states is non-zero but lower than the density of states
for the free electron system.
\vspace{0.5cm}

{\bf{ B. The two-dimensional system}}
\vspace{0.5cm}

In this case the Fourier form of the interacting potential is:
$$V_{\vec{p}-\vec{k}}=\int d^{2}r\;V(r) e^{-i(\vec{p}-\vec{k})\vec{r}}\eqno{(53)}$$
and we will use the $2D$ geometry, $\theta$ being the angle between $\vec{p}$ and $\vec{r}$, $\theta '$ the angle between $\vec{p}$ and $\vec{k}$,
to obtain:
$$V_{\vec{p}-\vec{k}}=\int_{0}^{\infty} dr\;r V(r) \int_{0}^{2\pi} d\theta\;e^{-i p r\cos\theta} e^{i k r \cos\alpha}\eqno{(54)}$$
with $\alpha=\theta-\theta '$. The angular integral $I$ will be evaluated using the expression of the exponential function with the Bessel functions [19]:
$$e^{i z\cos\varphi}=J_{0}(z)+2\sum_{n=1}^{\infty} i^{n} J_{n}(z)\cos(n\varphi)\eqno{(55)}$$
to obtain:
$$I=2\pi\sum_{l=0}^{\infty}\varepsilon_{l} J_{l}(p r)J_{l}(k r)\cos\left(l \theta' \right)\eqno{(56)}$$
with $\varepsilon_{l}=1$ for $l=0$, and $\varepsilon_{l}=2$ for $l=1,2,3,...$. Finally, $V_{\vec{p}-\vec{k}}$ will be:
$$V_{\vec{p}-\vec{k}}=\sum_{l=0}^{\infty}V_{l}(p, k)\cos\left(l \theta '\right)\eqno{(57)}$$
where:
$$V_{l}(p,k)=2\pi\varepsilon_{l}\int_{0}^{\infty}dr\; r V(r)\;J_{l}(p r) J_{l}(k r)\eqno{(58)}$$
The Hartree-Fock energy, in two dimensions, will be:
$$\varepsilon_{p}=
\frac{p^{2}}{2m}-\int_{0}^{p_{F}}\int_{0}^{2\pi}\frac{k dk d\theta'}{(2\pi)^{2}}\sum_{l=0}^{\infty} V_{l}(p,k)\cos\left(l\theta'\right)\eqno{(59)}$$
and it is easy transformed to:
$$\varepsilon_{p}=\frac{p^{2}}{2m}-\int_{0}^{p_{F}}k\;dk\int_{0}^{\infty}dr\;r V(r) J_{0}(p r) J_{0}(k r)\eqno{(60)}$$
Using now the non-Coulomb form of the interaction potential (16), after performing the $k$-integration, and using again the notations
$y=p/p_{F}$ and $z=p_{F} r$, we obtain:
$$\varepsilon_{p}=\frac{p^{2}}{2m}-A p_{F}^{\eta}\int_{0}^{\infty}dz\;z^{-\eta} J_{0}(y z) J_{1}(z)\eqno{(61)}$$
and the result, after $z$-integration, is:
$$\varepsilon_{p}=\frac{p^{2}}{2m}-A p_{F}^{\eta}
\left\{\begin{array}{lll}
\frac{\Gamma\left(1-\frac{\eta}{2}\right)}{2^{\eta}\Gamma\left(1+\frac{\eta}{2}\right)}F\left(1-\frac{\eta}{2}, -\frac{\eta}{2} ; 1 ; y^{2}\right) & \mbox{; $y<1$} \\
\frac{\Gamma(\eta)\Gamma\left(1-\frac{\eta}{2}\right)}{2^{\eta}\Gamma\left(\frac{\eta}{2}\right)\left[\Gamma\left(1+\frac{\eta}{2}\right)\right]^{2}} & \mbox{; $y=1$} \\
\frac{\Gamma\left(1-\frac{\eta}{2}\right)}{2^{\eta}y^{2-\eta} \Gamma\left(\frac{\eta}{2}\right)}F\left(1-\frac{\eta}{2}, 1-\frac{\eta}{2} ; 2 ; \frac{1}{y^{2}}\right) & \mbox{; $y>1$}
\end{array}\right.\eqno{(62)}$$
for $0<\eta<2$.
As in the three-dimensional case, in two dimensions and at small values of the momentum, $\varepsilon_{p}$ is negative due to the exchange hole space around electrons.
This behavior is maintained until around the Fermi momentum, for $\eta\le 1$.
Close to the Fermi momentum, for $p\rightarrow p_{F}$, $p<p_{F}$, following similar steps as for the $3D$ case, we obtain:
$$\varepsilon_{p}\simeq\frac{p_{F}^{2}}{2m}-A p_{F}^{\eta}\frac{\Gamma\left(1-\frac{\eta}{2}\right)}{2^{\eta}\Gamma\left(1+\frac{\eta}{2}\right)}\times $$
$$\times \left\{\frac{\Gamma(\eta)}{\Gamma\left(\frac{\eta}{2}\right)\Gamma\left(1+\frac{\eta}{2}\right)}{\cal{G}}_{1}+
\left(1-y^{2}\right)^{\eta} \frac{\Gamma(-\eta)}{\Gamma\left(1-\frac{\eta}{2}\right)\Gamma\left(-\frac{\eta}{2}\right)}{\cal{G}}_{2}\right\}\eqno{(63)}$$
where:
$${\cal{G}}_{1,2}\simeq 1\mp \frac{\eta (2\mp \eta)}{4 (1\mp \eta)}\left(1-y^{2}\right)\eqno{(64)}$$
For $p\rightarrow p_{F}$, $p>p_{F}$ the result will be:
$$\varepsilon_{p}\simeq\frac{p_{F}^{2}}{2m}-A p_{F}^{\eta}\frac{\Gamma\left(1-\frac{\eta}{2}\right)}{2^{\eta}y^{2-\eta} \Gamma\left(\frac{\eta}{2}\right)}\times $$
$$\times \left\{\frac{\Gamma(\eta)}{\left[\Gamma\left(1+\frac{\eta}{2}\right)\right]^{2}}{\cal{G}}_{3}+
\left(1-\frac{1}{y^{2}}\right)^{\eta} \frac{\Gamma(-\eta)}{\left[\Gamma\left(1-\frac{\eta}{2}\right)\right]^{2}}{\cal{G}}_{4}\right\}\eqno{(65)}$$
with:
$${\cal{G}}_{3,4}\simeq 1+ \frac{(2\mp \eta)^{2}}{4 (1\mp \eta)}\left(1-\frac{1}{y^{2}}\right)\eqno{(66)}$$
From eq.(62) the Coulomb result, obtained taking again $A=e^{2}$ and $\eta=1$, is:
$$\varepsilon_{p}=\frac{p^{2}}{2m}-\frac{2 e^{2} p_{F}}{\pi}
\left\{\begin{array}{lll}
\frac{\pi}{2}\;F\left(\frac{1}{2}, -\frac{1}{2} ; 1 ; y^{2}\right) & \mbox{; $y<1$} \\
1 & \mbox{; $y=1$} \\
\frac{\pi}{4 y}\;F\left(\frac{1}{2}, \frac{1}{2} ; 2 ; \frac{1}{y^{2}}\right) & \mbox{; $y>1$}
\end{array}\right.\eqno{(67)}$$
Using the following identities:
$$\frac{\pi}{2}\;F\left(\frac{1}{2}, -\frac{1}{2} ; 1 ; y^{2}\right)=E(y)\eqno{(68)}$$
and:
$$F\left(\frac{1}{2}, \frac{1}{2} ; 2 ; \frac{1}{y^{2}}\right)=\frac{4 y^{2}}{\pi}\left[E\left(\frac{1}{y}\right)-\left(1-\frac{1}{y^{2}}\right) K\left(\frac{1}{y}\right)\right]\eqno{(69)}$$
where $E(y)$ is the complete elliptic integral of the second kind, and $K(y)$ is the complete elliptic integral of the first kind, one obtains the classical result [5, 25]:
$$\varepsilon_{p}=\frac{p^{2}}{2m}-\frac{2 e^{2} p_{F}}{\pi}
\left\{\begin{array}{lll}
E(y) & \mbox{; $y<1$} \\
1 & \mbox{; $y=1$} \\
y \left[E\left(\frac{1}{y}\right)-\left(1-\frac{1}{y^{2}}\right) K\left(\frac{1}{y}\right)\right] & \mbox{; $y>1$}
\end{array}\right.\eqno{(70)}$$
The $2D$ Hartree-Fock energy for $N$ particles is given by:
$$E_{HF}=E_{HF}^{(1)}+E_{HF}^{(2)}\eqno{(71)}$$
where, in $2D$:
$$E_{HF}^{(1)}=\frac{1}{2}\;\frac{p_{F}^{2}}{2m}\;N\eqno{(72)}$$
and:
$$E_{HF}^{(2)}=-\frac{A p_{F}^{\eta+2}\Gamma\left(1-\frac{\eta}{2}\right)\; Y_{\eta}}{2^{\eta+1}\pi\Gamma\left(1+\frac{\eta}{2}\right)}\eqno{(73)}$$
with:
$$Y_{\eta}=\int_{0}^{1}dy\;y F\left(1-\frac{\eta}{2}, -\frac{\eta}{2}; 1; y^{2}\right)\eqno{(74)}$$
For the Coulomb case $Y_{\eta=1}=4/(3\pi)$, and:
$$E_{HF}^{(2)}=-\frac{2 e^{2}p_{F}^{3}}{3\pi^{2}}\eqno{(75)}$$
For the general case, using the new variable $x=y^{2}$ in eq.(74), and the integral (33), the scaled Hartree-Fock energy in two dimensions becomes:
$$\frac{E_{HF}}{N}=\frac{1}{2}\;\frac{p_{F}^{2}}{2m}-A\left(\frac{p_{F}}{2}\right)^{\eta}\frac{\Gamma\left(1-\frac{\eta}{2}\right)\Gamma(\eta)}
{\left(1+\frac{\eta}{2}\right)\left(\frac{\eta}{2}\right)^{2}\left[\Gamma\left(\frac{\eta}{2}\right)\right]^{3}}\eqno{(76)}$$
The velocity of the particle, for $y=p/p_{F}<1$, is easy obtained as:
$$v_{p}=\frac{p}{m}+A p_{F}^{\eta-2}\frac{\eta\Gamma\left(2-\frac{\eta}{2}\right)}{2^{\eta}\Gamma\left(1+\frac{\eta}{2}\right)}\;p\;F\left(2-\frac{\eta}{2}, 1-\frac{\eta}{2}; 2;
\frac{p^{2}}{p_{F}^{2}}\right)\eqno{(77)}$$
For the Coulomb case it reduces to:
$$v_{p}=\frac{p}{m}+\frac{e^{2}}{2}\;y\;F\left(\frac{3}{2}, \frac{1}{2}; 2; y^{2}\right)\eqno{(78)}$$
With eq.(37), and using:
$$\frac{\pi}{2}\;F\left(\frac{1}{2}, \frac{1}{2}; 1; y^{2}\right)=K(y)\eqno{(79)}$$
$v_{p}$ is rewritten as:
$$v_{p}=\frac{p}{m}+\frac{2 e^{2}}{\pi y}\left[K(y)-E(y)\right]\eqno{(80)}$$
For the case $y=p/p_{F}>1$ the velocity will be:
$$v_{p}=\frac{p}{m}+2 A p_{F}^{\eta-1}\frac{\Gamma\left(2-\frac{\eta}{2}\right)}{2^{\eta} \Gamma\left(\frac{\eta}{2}\right)}
\left(\frac{p_{F}}{p}\right)^{3-\eta}\times $$
$$\times\left\{F\left(1-\frac{\eta}{2}, 1-\frac{\eta}{2}; 2; \frac{p_{F}^{2}}{p^{2}}\right)+\frac{1}{2}\left(1-\frac{\eta}{2}\right)\left(\frac{p_{F}}{p}\right)^{2}
F\left(2-\frac{\eta}{2}, 2-\frac{\eta}{2}; 3; \frac{p_{F}^{2}}{p^{2}}\right)\right\}\eqno{(81)}$$
For the Coulomb case, using again eq.(37), the velocity will be:
$$v_{p}=\frac{p}{m}+\frac{2 e^{2}}{\pi}\left[K\left(\frac{1}{y}\right)-E\left(\frac{1}{y}\right)\right]\eqno{(82)}$$
Both velocities, in the Coulomb case, diverges at $p=p_{F}$.
Now we will approximate the velocities for $p$ close to the Fermi momentum $p_{F}$. In a similar way as in $3D$ case, for $p\rightarrow p_{F}$, $p<p_{F}$, we obtain:
$$v_{p\rightarrow p_{F}}\simeq\frac{p_{F}}{m}+ A p_{F}^{\eta-1}\frac{\eta \Gamma\left(2-\frac{\eta}{2}\right)}{2^{\eta} \Gamma\left(1+\frac{\eta}{2}\right)}\times $$
$$\times\left\{\frac{\Gamma(-1+\eta)}{\Gamma\left(\frac{\eta}{2}\right)\Gamma\left(1+\frac{\eta}{2}\right)}{\cal{T}}_{1}+
\left(1-y^{2}\right)^{-1+\eta}\frac{\Gamma(1-\eta)}{\Gamma\left(2-\frac{\eta}{2}\right)\Gamma\left(1-\frac{\eta}{2}\right)}{\cal{T}}_{2}\right\}\eqno{(83)}$$
where:
$${\cal{T}}_{1}\simeq 1+\frac{4-\eta}{4}\left(1-y^{2}\right)\eqno{(84)}$$
and:
$${\cal{T}}_{2}\simeq 1+\frac{2+\eta}{4}\left(1-y^{2}\right)\eqno{(85)}$$
For $\eta>1$, at the Fermi momentum, the velocity is given by:
$$v_{p=p_{F}}=\frac{p_{F}}{2}\left\{\frac{2}{m}+A\left(\frac{p_{F}}{2}\right)^{\eta-2}\frac{2}{\eta}\;\frac{\Gamma\left(2-\frac{\eta}{2}\right)
\Gamma(-1+\eta)}{\left[\Gamma\left(\frac{\eta}{2}\right)\right]^{3}}\right\}\eqno{(86)}$$
For $\eta<1$, keeping the dominant term, close to the Fermi momentum, the velocity will be:
$$v_{p\rightarrow p_{F}}\simeq\frac{A\Gamma(1-\eta)}{\pi}\left(\frac{p_{F}}{2}\right)^{\eta-1}\left(1-\frac{p^{2}}{p_{F}^{2}}\right)^{-1+\eta}\sin\left(\frac{\pi\eta}{2}\right)\eqno{(87)}$$
On the opposite side, for $p\rightarrow p_{F}$, $p>p_{F}$, the velocity is approximated as:
$$v_{p\rightarrow p_{F}}\simeq\frac{p_{F}}{m}+2A p_{F}^{\eta-1}\frac{\Gamma\left(2-\frac{\eta}{2}\right)}{2^{\eta}\Gamma\left(\frac{\eta}{2}\right)}\times$$
$$\times\left\{\frac{\Gamma(\eta)}{\left[\Gamma\left(1+\frac{\eta}{2}\right)\right]^{2}}{\cal{Z}}_{1}+
\left(1-\frac{1}{y^{2}}\right)^{\eta}\frac{\Gamma(-\eta)}{\left[\Gamma\left(1-\frac{\eta}{2}\right)\right]^{2}}{\cal{Z}}_{2}+\right.$$
$$\left.+\frac{1}{2}\left(1-\frac{\eta}{2}\right)\left[\frac{2 \Gamma(-1+\eta)}{\left[\Gamma\left(1+\frac{\eta}{2}\right)\right]^{2}}{\cal{H}}_{1}+
\left(1-\frac{1}{y^{2}}\right)^{-1+\eta}\frac{2 \Gamma(1-\eta)}{\left[\Gamma\left(2-\frac{\eta}{2}\right)\right]^{2}}{\cal{H}}_{2}\right]\right\}
\eqno{(88)}$$
with:
$${\cal{Z}}_{1,2}\simeq 1+\frac{(2\mp\eta)^{2}}{4 (1\mp\eta)}\left(1-\frac{1}{y^{2}}\right)\eqno{(89)}$$
and:
$${\cal{H}}_{1}\simeq 1+\frac{(4-\eta)^{2}}{4 (2-\eta)}\left(1-\frac{1}{y^{2}}\right)\eqno{(90)}$$
$${\cal{H}}_{2}\simeq 1+\frac{(2+\eta)^{2}}{4\eta}\left(1-\frac{1}{y^{2}}\right)\eqno{(91)}$$
In this case, for $\eta>1$, and at the Fermi momentum, we reobtain the result given by eq.(86), and for $\eta<1$ the result is:
$$v_{p\rightarrow p_{F}}\simeq\frac{A\Gamma(1-\eta)}{\pi}\left(\frac{p_{F}}{2}\right)^{\eta-1}\left(1-\frac{p_{F}^{2}}{p^{2}}\right)^{-1+\eta}\sin\left(\frac{\pi\eta}{2}\right)\eqno{(92)}$$
Using these results one can calculate the $2D$ density of states $N(p)$ given by:
$$N(p)=\frac{p}{\pi \left|v_{p}\right|}\eqno{(93)}$$
For $\eta>1$ the non-zero density of states at the Fermi momentum is:
$$N(p_{F})=\frac{1}{\pi\left\{\frac{1}{m}+A\left(\frac{p_{F}}{2}\right)^{\eta-2}\frac{\Gamma\left(2-\frac{\eta}{2}\right)\Gamma(-1+\eta)}
{\eta\left[\Gamma\left(\frac{\eta}{2}\right)\right]^{3}}\right\}}\eqno{(94)}$$
For $\eta<1$, close to the Fermi momentum, the density of states is given by:
$$N\left(p\rightarrow p_{F}\right)\simeq\frac{2^{\eta-1}}{A \Gamma(1-\eta)\left|\sin\left(\frac{\pi\eta}{2}\right)\right|}\; p_{F}^{2-\eta}\times
\left\{\begin{array}{ll}
\left(1-\frac{p^{2}}{p_{F}^{2}}\right)^{1-\eta} & \mbox{; $p<p_{F}$} \\
\left(1-\frac{p_{F}^{2}}{p^{2}}\right)^{1-\eta} & \mbox{; $p>p_{F}$}
\end{array}\right.\eqno{(95)}$$
In the case of the sub-Coulomb interaction ($\eta>1$) the density of states at the Fermi momentum is non-zero and lower than the free electrons case.
On the other hand, in the case of the super-long type interaction ($\eta<1$), the density of states is zero at the Fermi momentum, as in the three-dimensional case,
the momentum dependence close to the Fermi momentum being also of the power function type but with a different coefficient.

\vspace{0.5cm}

\begin{large}
	{\bf III. Discussions}
\end{large}
\vspace{0.5cm}

In this section we discuss the exact analytical results obtained for a system of interacting fermions through a non-Coulomb type interaction.
The analysis was done for both three-dimensional and two-dimensional systems. The interaction between the particles was modeled by a distance dependent
potential of the form $r^{-\eta}$, for $\eta<1$ having the super-long range type potential, and for $\eta>1$ the sub-Coulomb type potential. For $\eta=1$ we have the
Coulomb case , and the well-known results from this case were reobtained, as particular cases, for both the three-dimensional and for the two-dimensional case.
We established general relations for the energy of the particle in the presence of interaction, for the Hartree-Fock ground states energy of the system, as well as
for the momentum dependent density of states. The density of states, which is a central quantity in statistical physics and in condensed matter, was determined as a function
of momentum and then its expression was approximated in the vicinity of the Fermi momentum. In the case of the super-long range type interaction the density of states is zero
at the Fermi momentum in both the three and two dimensions. Unlike the Coulomb case, where in the vicinity of the Fermi momentum  the momentum dependence of the density of states
is logarithmic, in the non-Coulomb case, for $\eta<1$, it is of the power function type. However, even in this case the results are unrealistic as long as the screening effects of
the interaction are neglected. On the other hand, the sub-Coulomb case ($\eta>1$), where the interactions have a shorter range of action, approaches (but not as an analytical expression)
to the case of screened interaction. In this case the density of states at the Fermi momentum is non-zero but lower than the free particles case.
 However, the results
obtained in this paper, even if they are somewhat unrealistic without screening, they are an extension of the well-known Coulomb case, and could be a platform for future investigations, such
as plasmon collective oscillations in condensed matter [26-34], Friedel oscillations [35-41], and other important properties from condensed matter physics.

\vspace{0.5cm}

\begin{large}
	{\bf  References}
\end{large}
\vspace{0.5cm}

[1] S.M.Girvin, K.Yang, \emph{Modern Condensed Matter Physics}, Cambridge Univ.

Press, (2019)

[2] M.L.Cohen, S.G.Louie, \emph{Fundamentals of Condensed Matter Physics},

Cambridge Univ.Press, (2016)

[3] D.Pines, P.Nozieres, \emph{The Theory of Quantum Liquids}, W.A.Benjamin,

New York, (1966)

[4] P.Phillips, \emph{Advanced Solid State Physics}, Cambridge Univ.Press,
(2014)

[5] G.F.Giuliani, G.Vignale, \emph{Quantum Theory of the Electron Liquid},

Cambridge Univ.Press, (2005)

[6] I.Ia.Pomeranchuk, Soviet Physics JETP 35, 524, (1958)

[7] V.A.Khodel, V.R.Shaginyan, V.V.Khodel, Physics Reports 249, 1, (1994)

[8] J.Quintanilla, M.Haque, A.J.Schofield, Phys.Rev.B 78, 035131, (2008)

[9] O.Ciftja, Physica B 458, 92, (2015)

[10] S.Hong, G.D.Mahan, Phys.Rev.B 52, 7860, (1995)

[11] N.Defenu, A.Lerose, S.Pappalardi, Phys.Reports 1074, 1, (2024)

[12] L.Radzihovski, J.Toner, arXiv:2401.04761

[13] L.N.Carenza, J-M.Armengol-Collado, D.Krommydas, L.Giomi,

arXiv:2311.03276

[14] A.M.Gabovich, M.S.Li, H.Szymczak, A.I.Voitenko, Journal of Electrostatics

102, 103377, (2019)

[15] J.Stafiej, D.di Caprio, J.P.Badiali, J.Chem.Phys.109, 3607, (1998)

[16] L.Zhang, H.Wang, M.C.Muniz, A.Z.Panagiotopoulas, R.Car, E.Weinan,

J.Chem.Phys.156, 124107, (2022)

[17] R.Vangara, K.Stoltzfus, M.R.York, F.van Swol, D.P.Petsev, Mater.Res.

Express 6, 086331, (2019)

[18] P.A.Bares, X.-G.Wen, Phys.Rev.B 48, 8636, (1993)

[19] A.Campa, T.Dauxois, S.Ruffo, Phys.Reports 480, 57, (2009)

[20] N.Defenu, T.Donner, T.Macri, G.Pagano, S.Ruffo, A.Trombettoni,

Rev.Mod.Phys.95, 035002, (2023)

[21] D.Laghi, T.Macri, A.Trombettoni, Phys.Rev.A 96, 043605, (2017)

[22] P.R.Wallace, \emph{Mathematical Analysis of Physical Problems}, Dover,

New York, (1984)

[23] I.S.Gradshteyn, I.M.Ryzhik, \emph{Table of Integrals, Series, and Products},

Academic Press, (1980)

[24] P.C.Consul, Bulletins de l'Academie Royale Belgique 52, 562, (1966)

[25] A.V.Chaplik, Soviet Physics JETP 33, 997, (1971)

[26] J.Zhang, L.Zhang, W.Xu, J.Phys.D: Appl.Phys.45, 113001, (2012)

[27] I.Grosu, L.Tugulan, Physics E 40, 474, (2008)

[28] O.Pluchery, J.-F.Bryche, \emph{An Introduction to Plasmonics},
World

Scientific, (2023)

[29] S.Das Sarma, W.-Y.Lai, Phys.Rev.B 32, 1401(R), (1985)

[30] S.Das Sarma, E.H.Hwang, Phys.Rev.B 54, 1936, (1996)

[31] B.Diaconescu, K.Pohl, L.Vattieone, L.Savio, P.Hofmann, V.M.Silkin,

J.M.Pitarke, E.V.Chulkov, P.M.Echenique, D.Farias, M.Rocca,

Nature 448, 57, (2007)

[32] J.M.Pitarke, V.M.Silkin, E.V.Chulkov, P.M.Echenique, Rep.Prog.Phys.70,

 1, (2007)

[33] M.S.Tame, K.R.McEnery, \c{S}.K.\"{O}zdemir, J.Lee, S.A.Maier, M.S.Kim,

Nature Physics 9, 329, (2013)

[34] E.H.Hwang, R.E.Throckmorton, S.Das Sarma, Phys.Rev.B 98, 195140, (2018)

[35] G.E.Simion, G.F.Giuliani, Phys.Rev.B 72, 045127, (2005)

[36] I.Grosu, L.Tugulan, J.Supercond.Nov.Magn.21, 65, (2008)

[37] C.Bena, Comptes Rendus Physique 17, 302, (2016)

[38] C.-Y.Lin, C.-W.Chiu, Scientific Reports 14, 13792, (2024)

[39] C.Dutreix, M.I.Katsnelson, Phys.Rev.B 93, 035413, (2016)

[40] T.Farajallahpour, S.Khamouei, S.S.Shateri, A.Phirouznia, Scientific

Reports 8, 2667, (2018)

[41] E.G.Dalla Tore, D.Benjamin, Y.He, D.Dentelski, E.Demler, Phys.Rev.B 93,

205117, (2016)

\end{document}